\documentclass[preprint,aps,nofootinbib]{revtex4}

\usepackage{graphicx}
\usepackage{dcolumn}
\usepackage{bm}
\usepackage{latexsym}
\usepackage{amsmath}
\usepackage{amsfonts}
\usepackage{units}
\usepackage{pifont}
\usepackage{cancel}
\usepackage{amssymb}
\usepackage{subfigure}
\usepackage{epsfig}

\long\def\symbolfootnote[#1]#2{\begingroup%
\def\thefootnote{\fnsymbol{footnote}}\footnote[#1]{#2}\endgroup}

\newcommand{\beann}{\begin{eqnarray*}}
\newcommand{\eeann}{\end{eqnarray*}}

\newcommand{\benn}{\begin{equation*}}
\newcommand{\eenn}{\end{equation*}}

\newcommand{\w}{\omega}
\newcommand{\D}{\delta}
\newcommand{\cosp}{\cos\frac{\Pi}{f_{\pi}}}
\newcommand{\G}{\gamma}

\newcommand{\ric}{\mathcal{R}}
\newcommand{\fpi}{f_{\pi}}

\newcommand{\aon}{\alpha_1}
\newcommand{\atw}{\alpha_2}
\newcommand{\Zed}{\mathbb{Z}}

\begin{document}

\title{Warped Hybrid Inflation}
\author{Raman Sundrum}
\email[]{sundrum@pha.jhu.edu}
\author{Christopher M. Wells}%
\email[]{cwells13@pha.jhu.edu}
\affiliation{
Department of Physics and Astronomy\\Johns Hopkins University\\3400 
North Charles Street\\Baltimore, MD 21218-2686
}

\date{\today}
\begin{abstract}
We construct a model of hybrid inflation within a  controlled five-dimensional
effective field theory framework.  The inflaton and waterfall fields are realized as naturally light
moduli of the 5D compactification. At the quantum level, waterfall loops must be cut off at a scale
considerably lower than the inflaton field transit in order to preserve  slow-roll dynamics without
fine-tuning.  We accomplish this by a significant warping, or redshift, between the extra-dimensional regions
in which the inflaton and waterfall fields are localized. The mechanisms we employ have been
separately realized in string theory, which suggests that a string UV completion of our model is possible.
We study a part of the parameter space in which the cosmology takes a standard form,
but we point out that it is also possible for some regions of space to end inflation by
quantum tunneling. Such regions may provide new cosmological signals, which we will study
in future work.

\end{abstract}

\maketitle

\section{Introduction}

\begin{figure}[h]
\centering
\includegraphics[scale=0.4]{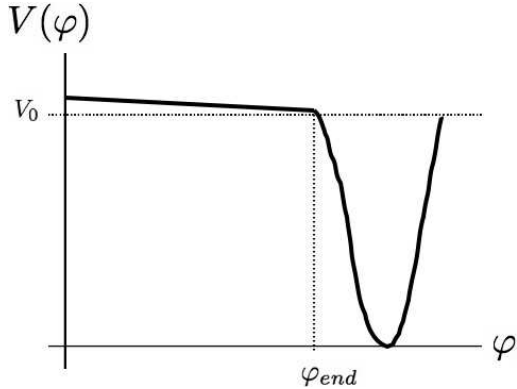}
\caption{The prototypical inflaton potential.  Achieving a flat potential which suddenly drops at $\varphi_{end}$ requires tuning the coefficients of a series of higher-dimensional operators.}
\label{fig:protoinflaton}
\end{figure}

The theory of Inflation provides an attractive approach to understanding 
the cosmological
initial conditions of our universe \cite{Guth:1980zm,Linde:1981mu,Baumann:2009ds}.
But the microscopic 
basis for inflation is highly challenging, given what we know and trust within
effective quantum field theory. For example, in the case of single-field 
inflation, in which the inflaton field remains safely sub-Planckian, the inflaton potential must have an approximately flat 
slow-roll regime followed by a rapid drop for re-heating (Fig.\,\ref{fig:protoinflaton}). 
It typically requires considerable 
tuning among different couplings. 

This raises the question of whether inflation in our universe 
is the result of a ``chance'' tuning among couplings or of a deeper 
mechanism. Directly or indirectly, 
the answer can have  observable ramifications. For example, 
the degree of tuning increases as the scale of inflation is lowered, 
so it is much more likely that an inflationary potential arises 
by chance for the case of high-scale inflation than for low-scale inflation.
On the other hand, some models of fundamental physics produce
 unwanted particles (for example, late-decaying gravitinos \cite{Krauss1983556}), defects, or other inhomogeneities
 at relatively low scales, 
features that can be erased by low-scale inflation.  
An underlying mechanism for  inflation 
 might have less of a high-scale bias and make such 
models more plausible. Of course, a particular mechanism for inflation may give rise to 
new cosmological signatures.

Several broad mechanisms have been discussed in the literature.
It seems natural to turn to symmetries in order to explain the special 
features of an inflationary potential, in particular the slow-roll region. 
Supersymmetric theories do often have ``flat directions'' in field space, 
but inflationary curvature breaks supersymmetry badly enough that it cannot
by itself explain the flatness of the inflaton potential \cite{Copeland:1994vg,Lyth:1998xn,Randall:1997kx}.
 Instead, in Natural Inflation \cite{Freese:1990rb} this flatness is ensured by realizing
 the inflaton as a  pseudo Nambu-Goldstone 
boson (PNGB), protected by the associated  (approximate) shift-symmetry.
But this still leaves the challenge of escaping from slow-roll to reheating. 
Hybrid Inflation \cite{Linde:1993cn} provides an elegant answer to the sudden change in shape 
of the 
potential. The central insight is that with multiple fields, the path in 
field space can take sudden turns  in response to even a gently varying 
potential landscape. 
 The sudden change in potential of Fig.\,\ref{fig:protoinflaton} can then be 
 traced out along such a ``bent'' path. 

Let us examine hybrid inflation more carefully, along the lines of Ref.\,\cite{ArkaniHamed:2003mz}.  The prototypical model of hybrid inflation has two fields, the inflaton field
$\Pi(x)$ and a ``waterfall'' field $\omega(x)$, with potential 
(in Einstein frame),
\begin{eqnarray}
\label{eq:protohybrid}
V &=& \frac{m_{\Pi}^2}{2} \Pi^2 + g_1 \Pi^2 \omega^2 + g_2 (\omega^2 - \omega_1^2)^2
\nonumber \\
&=&  \frac{m_{\Pi}^2}{2} \Pi^2 + (g_1 \Pi^2 - 2 g_2 \omega_1^2) \omega^2 +
g_2 \omega^4 + g_2 \omega_1^4,
\end{eqnarray}
where $\omega_1$ sets the true vacuum, $\omega = \omega_1, \Pi = 0$ at zero 
vacuum energy. 
However if one starts cosmic evolution at sufficiently large 
$\Pi = \Pi_0 \neq 0$, then 
the second line shows that one can effectively have a $\Pi$-dependent
mass-squared for $\omega$, which stabilizes $\omega =0$. Plugging this back 
into $V$, gives an effective potential for just $\Pi$, 
\begin{equation}
V_{eff}(\Pi) = \frac{m_{\Pi}^2}{2} \Pi^2 + g_2 \omega_1^4.
\end{equation}
For small $m_{\Pi}^2$, $\Pi$ rolls slowly to smaller values 
and results in inflation, with Hubble constant given by 
$H^2 M_{Pl}^2 \simeq g_2 \omega_1^4$. 
While $V_{eff}$ itself does not describe how inflation 
ends, it does end suddenly when the $\Pi$-dependent mass term for $\omega$ 
turns tachyonic, and $\omega$ is destabilized from the origin towards the 
true vacuum at $\omega_1$. 

It has been suggested that 
the  smallness of 
$m_{\Pi}^2$ can be explained by realizing $\Pi$ as a PNGB \cite{ArkaniHamed:2003mz,Kaplan:2003aj}. This is 
an attractive strategy, using approximate shift symmetry to do what it 
does best,
protecting  the flatness of the inflaton potential, while using the 
waterfall mechanism to do what it does best, namely  ending inflation.
But at the quantum level we must check that the levels of  
shift-symmetry breaking by $m_{\Pi}^2$ and $g_1$ are naturally compatible.
Indeed, without further structure they are not, as can be seen by studying the 
one-loop renormalization of $m_{\Pi}^2$ by $g_1$, 
\begin{equation}
\Delta_{\omega-loop} m_{\Pi}^2 \sim \frac{g_1 \Lambda_{UV}^2}{16 \pi^2},
\end{equation}
where $\Lambda_{UV}$ is the cutoff of this effective field theory.

At the start of inflation we need positive $\Pi$-dependent mass-squared for 
$\omega$,
\begin{equation}
g_1 \Pi_0^2 > 2 g_2 \omega_1^2 \simeq \frac{H^2 M_{Pl}^2}{\omega_1^2}, 
\end{equation} 
which then implies, 
\begin{equation}
\Delta_{\omega-loop} m_{\Pi}^2 \gtrsim 
\frac{M_{Pl}^2 \Lambda_{UV}^2}{16 \pi^2 \omega_1^2 \Pi_0^2} H^2,
\end{equation}
Technical naturalness requires $m_{\Pi}^2 \gtrsim
 \Delta_{\omega-loop} m_{\Pi}^2$, 
while the slow-roll conditions require $m_{\Pi}^2 \ll H^2$, from which we 
conclude that 
\begin{equation} 
\label{eq:cutoffratio}
\frac{M_{Pl} \Lambda_{UV}}{\omega_1 \Pi_0} < 1.
\end{equation} 
But  this is only possible for fields larger than the cutoff or  the 
Planck scale. Moving away from this dangerous regime for effective
field theory, the model rapidly becomes fine-tuned. 
  
Refs.\,\cite{ArkaniHamed:2003mz,Kaplan:2003aj} describe new physics that can cut off the quadratic divergence above 
and thereby resolve the fine-tuning problem, either based on extra dimensions, supersymmetry or 
the Little Higgs mechanism. In the present paper we describe a new approach 
that can be thought of as using compositeness of the inflaton and waterfall 
fields in order to cut off  UV sensitivity. If this is the case, 
one would expect the compositeness scale to cut off quantum loops involving the
light composite fields. But this would appear to only reinterpret 
$\Lambda_{UV}$ as a physical compositeness scale, 
without altering the conclusion that fields must take on Planckian values 
in order to preserve naturalness. We therefore take
 $\Pi$ and $\omega$ 
to be composites of two  separate sectors, 
 with  {\it different} compositeness
scales, $\Lambda_{\Pi}$ and $\Lambda_{\omega}$ respectively. 
The $\omega$-loop would then be cut off by $\Lambda_{UV} \equiv 
\Lambda_{\omega}$. Naturalness can then be satisfied with sub-Planckian and 
sub-cutoff fields if 
\begin{equation}
\Lambda_{\Pi} \gg \Lambda_{\omega}.
\end{equation}
In particular, Eq.\,(\ref{eq:cutoffratio}) can be satisfied by  $\Lambda_{UV} \equiv 
\Lambda_{\omega} \ll \Pi_0$.

A theory based on compositeness would ordinarily run into the requirement of 
understanding the underlying strong coupling dynamics. However, 
exploiting AdS/CFT duality \cite{Maldacena:1997re,Gubser:1998bc,Aharony:1999ti,Witten:1998qj} we can present the same ideas 
in terms of weakly-coupled higher-dimensional (minimally 5D) 
effective field theory 
within warped throats, one for each of the two composite sectors.  Here we proceed minimally within 5D effective field theory based on the Randall-Sundrum I (RS1) model \cite{Randall:1999ee}. The connections between such throats and 4D CFT's were developed in Ref's \,\cite{Verlinde:1999fy,maldacenaunpub,wittenITP,Verlinde:1999xm,Gubser:1999vj,ArkaniHamed:2000ds,Rattazzi:2000hs,PerezVictoria:2001pa}
The fields $\Pi$ and $\omega$ are then realized as light moduli within 
each of the two throats. It is attractive to realize $\Pi$ as a composite 
PNGB for the reasons described above. The AdS/CFT dual of this is 
that $\Pi$ is the fifth component of a 5D gauge field \cite{Contino:2003ve}. 
In the next section we will argue for identifying the waterfall 
field $\omega$ with the ``radion'' modulus of an RS1 throat.

With such an identification, the relaxation of
$\omega$  from a metastable VEV to its true-vacuum VEV 
corresponds to the motion of the IR boundary of the RS1 throat. 
In this sense it superficially 
resembles the approach of Brane Inflation \cite{Dvali:1998pa,Kachru:2003sx}. 
The difference is that in Brane Inflation it is the inflaton that is 
realized as a mobile brane while here it is the waterfall field that is a 
mobile ``brane''. The usual Brane Inflation strategy is to 
geometrically realize  inflaton shift-symmetry as approximate
 brane translation 
invariance when the inflaton brane is far from other higher-dimensional 
objects. However, this symmetry is typically spoiled once the higher dimensions
are compactified down to 4D, and gravitational backreaction taken into account \cite{Kachru:2003sx}.  By comparison, 
shift symmetry in extra-dimensional components of 
gauge fields is quite robust.

\begin{figure}[ht]
\centering
		
		\subfigure[\,Standard Hybrid Inflation]{
		\label{fig:animatestdhybrid}
		\includegraphics[width = 4.5 in, height = 1.75 in]{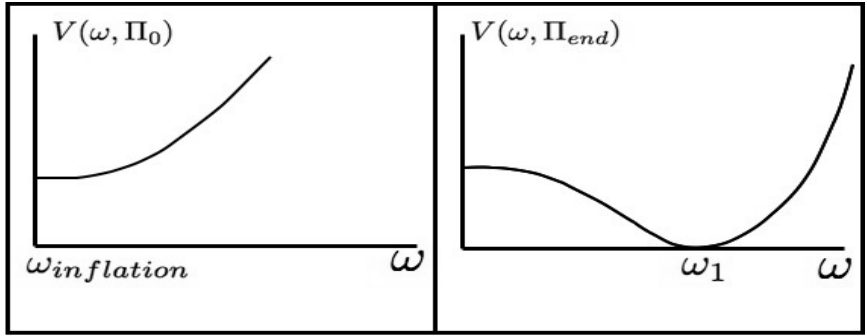}
		}
		\hspace{2 in}
	\subfigure[\,Warped Hybrid Inflation]{
		\label{fig:animateourhybrid}
		\includegraphics[width = 4.5 in, height = 1.75 in]{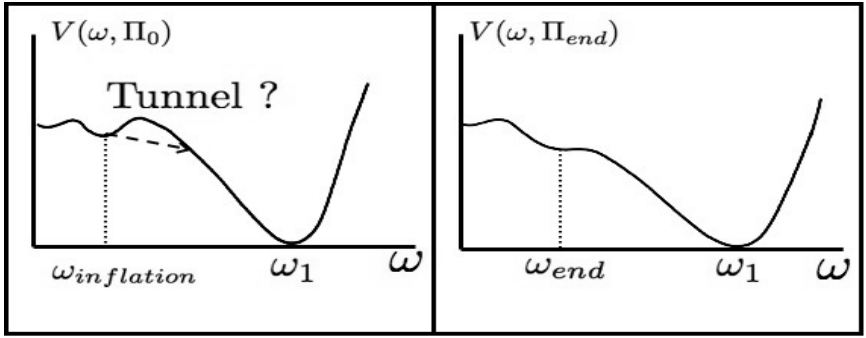}
		}
\caption{`Animations' of standard Hybrid Inflation - as in Eq.\,(\ref{eq:protohybrid}) - and Warped Hybrid Inflation.  Note that in the the warped case, inflation takes place while the waterfall is at a \textit{metastable} minimum.}
\label{fig:animatecomparison}
\end{figure}

There is an important difference between the effective potential of our 
model and the model of Eq.\,(\ref{eq:protohybrid}), illustrated in Fig.\,\ref{fig:animatecomparison}. 
Here we plot both potentials as a function of the waterfall field for 
the inflationary value of the inflaton. While in the model of elementary 
fields - see Fig.\,\ref{fig:animatestdhybrid} - the inflationary VEV of $\omega$ is the true minimum of the 
potential at $\Pi_0$, in our model - see Fig.\,\ref{fig:animateourhybrid} - the inflationary VEV is a metastable local minimum. 
In both cases as the inflaton rolls, the picture changes and the 
inflationary $\omega$ VEV is destabilized classically towards the true vacuum. 
But in our model, 
there is always the alternate possibility of quantum
 tunnelling to the true vacuum earlier. This would be a phenomenological 
disaster if tunneling dominated the end of inflation, but may 
provide an interesting phenomenology if it is subdominant \cite{Guth:1980zm,Liddle:1991tr}. In this paper, 
we will study the issue of tunneling just 
enough to choose a region of parameter
space where it is completely suppressed in our universe. But in future 
work we will return to study its phenomenological prospects.  

While we present our model within 5D effective field theory, the ingredients
of the model seem to be compatible  with string theory, as it 
is currently understood, and we hope that our work will help identify 
inflating string theories within the ``landscape'' \cite{Susskind:2003kw}.  This raises the question
of the role of supersymmetry. In our model, supersymmetry plays almost no 
role, although it is compatible with high-scale supersymmetry. The possible 
auxiliary role will be commented on  in the last section. 
In this paper, we will 
focus on arriving at a natural theory of {\it high-scale}
 inflation, because this 
is the easiest target. However, our goals are broader and include understanding
whether low-scale inflation can naturally occur. We expect that 
supersymmetry plays a more important role here. But we will leave such 
an investigation for future work.

The paper is organized as follows. In Section \ref{sec:strategy}, we will 
motivate in the language of strongly-coupled 
compositeness, the strategy behind our 
model, and then translate the key ingredients into the language of 
weakly-coupled warped 5D effective field theory using the AdS/CFT duality.
In Section \ref{sec:5D} we present our two-throat 
model, their light moduli, and their couplings. Because 
$\Lambda_{\Pi}$ need not be too small we will realize the dual inflaton 
throat as negligibly warped.
In Section \ref{sec:4Dtwofield} we derive the low-energy 
effective $\Pi-\omega$ theory and a variety of constraints that simplify 
our analysis and ensure  a generalized hybrid inflation structure for the 
effective potential.  In Section \ref{sec:tunnel} we discuss 
quantum tunneling from the metastable to stable $\omega$ vacuum.
 In Section \ref{sec:onefield} we integrate out 
$\omega$ during inflation and arrive at an effective theory for just $\Pi$, 
and check the slow-roll conditions.  In Section \ref{sec:leadcorr} we estimate the leading quantum corrections to the 5D action.  In Section \ref{sec:sample} we present an illustrative 
set of parameters, which are not fine-tuned, in which successful inflation 
takes place. 
In Section \ref{sec:disc} we discuss our results.

\section{Compositeness Strategy}
\label{sec:strategy}

Let us begin by thinking of our two fields as composites
 of different strongly-interacting sectors in purely 4D spacetime. 
We ask how a ``waterfall'' cross-coupling of $\Pi$ to $\omega$ can 
arise. In general cross-couplings arise  by multiplying composite operators 
(which interpolate the composites) 
from each of the two sectors within the Planck-scale  Lagrangian,
\begin{equation}
{\cal L}(M_{Pl}) \ni
 \frac{{\cal O}_{\Pi}(x) {\cal O}_{\omega}(x)}{M_{Pl}^{d_{\Pi} + 
d_{\omega} -4}},
\end{equation}
where the operators have high-energy scaling dimensions $d_{\Pi}$ and 
$d_{\omega}$. Below $\Lambda_{\Pi}$, ${\cal O}_{\Pi}$ ``hadronizes'' 
into some function of the light composite $\Pi$,
\begin{equation}
{\cal L}_{eff}(\Lambda_{\Pi}) \ni
 \frac{\Lambda_{\Pi}^{d_{\Pi}} f(\Pi/\Lambda_{\Pi}) 
{\cal O}_{\omega}(x)}{M_{Pl}^{d_{\Pi} + d_{\omega} -4}}.
\end{equation}
The generalization  of a $\Pi$-dependent ``mass term'' for $\omega$, 
is a $\Pi$-dependent coefficient of a relevant operator in the 
$\omega$-sector, that is,
\begin{equation}
\label{eq:needtachyon}
d_{\omega} < 4.
\end{equation}
Below $\Lambda_{\omega}$ this sector also hadronizes and we obtain 
\begin{equation}
{\cal L}_{eff}(\Lambda_{\omega}) \ni
 \frac{\Lambda_{\Pi}^{d_{\Pi}} f(\Pi/\Lambda_{\Pi}) 
\Lambda_{\omega}^{d_{\omega}} g(\omega/\Lambda_{\omega})}{M_{Pl}^{d_{\Pi} + 
d_{\omega} -4}}.
\end{equation}

However, there is a problem with the general structure of such a 
cross-coupling for the purpose of inflation, namely 
$g(\omega_{inflation}/\Lambda_{\omega})$ has no particular reason to be small, 
which in turn leads to an inflaton mass contribution, 
\begin{equation}
\Delta m_{\Pi}^2 \sim \frac{\Lambda_{\Pi}^{d_{\Pi} -2} \Lambda_{\omega}^{d_{\omega}}}{M_{Pl}^{d_{\Pi} + 
d_{\omega} -4}}.
\end{equation}
We can estimate this by noting that we want different values of the 
 cross-coupling (for 
differing values of $\Pi$)
to be responsible for the 
drop in potential energy density, from inflation at $\sim H^2 M_{Pl}^2$, 
to the true vacuum at zero. Without any special structure,  
\begin{equation}
\frac{\Lambda_{\Pi}^{d_{\Pi}} \Lambda_{\omega}^{d_{\omega}}}{
M_{Pl}^{d_{\Pi} + d_{\omega} -4}} \gtrsim H^2 M_{Pl}^2.
\end{equation}
We therefore find,
\begin{equation}
\Delta m_{\Pi}^2 \gtrsim H^2 \frac{M_{Pl}^2}{\Lambda_{\Pi}^2}, 
\end{equation}
which violates slow-roll (since $\Lambda_{\Pi} \lesssim M_{Pl}$),
unless fine-tuned away. 
This should be compared with the model of 
Eq.\,(\ref{eq:protohybrid}), in which the cross-coupling itself (at $\Pi_0$) drives
$\omega \rightarrow 0$, which in turn suppresses the $\Pi$ mass 
contribution from this coupling.

But there is a special kind of strongly-interacting
 composite theory that shares this  feature of the elementary model,
namely a conformal field theory (CFT) which undergoes spontaneous 
breaking of conformal invariance. It must give rise to a (composite) 
Nambu-Goldstone boson, which we will identify with $\omega$.  In such a 
scale-invariant dynamics, there is no independent compositeness scale 
$\Lambda_{\omega}$, but instead
\begin{equation} 
\Lambda_{\omega} \equiv \langle \omega \rangle.
\end{equation} 
The IR cross-coupling then takes the form
\begin{equation}
{\cal L}_{eff}(\Lambda_{\omega}) \ni
 \frac{\Lambda_{\Pi}^{d_{\Pi}} f(\Pi/\Lambda_{\Pi}) 
\omega^{d_{\omega}}}{M_{Pl}^{d_{\Pi} + 
d_{\omega} -4}}.
\end{equation}
As in the elementary model,
 this potential itself can drive $\omega$ to small values during 
inflation and large values at reheating, but unlike the 
elementary model a small (but non-zero) $\omega_{inflation}$ also acts as
 a low cutoff on the $\omega$ loops renormalizing the inflaton mass.

A second issue is that given Eq.\,(\ref{eq:needtachyon}), 
 we must ask why our Lagrangian
 does not naturally contain 
\begin{equation}
{\cal L}(M_{Pl}) \ni M_{Pl}^{4 - d_{\omega}} {\cal O}_{\omega},
\end{equation}
which would overwhelm all other dynamics with its Planckian mass scale.
 In the model 
of Eq.\,(\ref{eq:protohybrid}), this is the question of why we did not write a mass term 
$M_{Pl}^2 \omega^2$ given we have $\Pi^2 \omega^2$. But in the case of 
compositeness such a relevant coupling can be forbidden if 
$ {\cal O}_{\Pi}$ and ${\cal O}_{\omega}$ transform under a high-energy symmetry, 
 such that 
only their product is invariant,
 for example if  they were each odd under a $\Zed_2$-symmetry. This symmetry 
can then be broken at the compositeness scales of each sector, in particular 
by $\langle \omega \rangle$ in the $\omega$ sector. 

Now let us use the AdS/CFT correspondence to translate the above elements 
to  weakly-coupled but higher-dimensional (minimally 5D) 
effective field theory 
within two warped throats, one for each of the two composite sectors. 
The fields $\Pi$ and $\omega$ are then realized as light moduli within 
each of these throats. It is attractive to realize $\Pi$ as a composite 
PNGB for the reasons described above. The AdS/CFT dual of this is 
to that $\Pi$ is the fifth component of a 5D gauge field. Taking 
$\omega$ to be the ``radion'' modulus of an RS1 throat 
gives the  minimal dual  
incarnation of a PNGB of spontaneous conformal symmetry breaking \cite{ArkaniHamed:2000ds,Rattazzi:2000hs}.
 The dual of the operators ${\cal O}$ are then scalar fields propagating 
within each throat, which can be coupled on their shared UV brane. 
The relevance of the operator ${\cal O}_{\omega}$ translates into 
it being an AdS tachyon in the RS1 throat. The danger this poses to the 
throat stability then requires assigning it symmetry quantum numbers, minimally
a discrete $Z_2$ symmetry. This summarizes the strategy underlying what 
follows.

\section{The Model}
\label{sec:5D}

\begin{figure}[ht]
\centering
\includegraphics[scale=0.5]{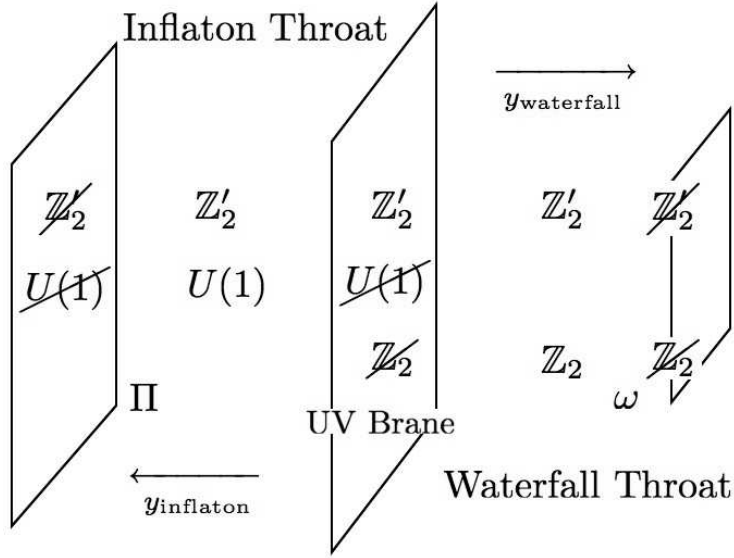}
\caption{Schematic for the bulk symmetries and brane localized symmetry breakings of the two throat system.  The $U(1)$ is gauged and its fifth component gives us a PNGB inflaton.  The parities $\mathbb{Z}_2$ and $\mathbb{Z}'_2$ are matter symmetries that constrain the symmetry breaking patterns in the waterfall throat.}
\label{fig:fieldsymm}
\end{figure}

The 5D model consists of an extra dimensional interval with two boundaries 
and one intermediate 3-brane  which splits spacetime into two regions, 
as shown in Fig.\,\ref{fig:fieldsymm}. One region we call the ``waterfall throat'', and it is 
a highly warped RS1-like throat with its light radion ultimately playing the 
role of the waterfall field of hybrid inflation. The intermediate brane acts 
as the UV brane for this throat. The second region is taken to be only 
mildly warped, although for convenience we will refer to it as the 
``inflaton throat.'' In it the 
inflaton of hybrid 
inflation is realized as the $A_5$ component of a 5D gauge 
field. With the exception of the 
metric of 5D General Relativity, 
each region has separate field content. Nevertheless, 
 non-gravitational fields can meet 
and interact at the intermediate UV brane. In particular, 
such a cross-coupling between throats will result in the 
coupling of inflaton and waterfall fields that plays a central 
role in hybrid inflation. 

The 5D action of the full 
theory takes the form
\begin{equation}
\label{eq:fullaction}
S = S_{infl} + S_{water} + S_{cross},
\end{equation}
where the first two terms on the right describe couplings in each of the 
two throats separately and the last term describes the origins of 
the inflaton-waterfall cross-coupling and related physics; see Figure \ref{fig:fieldsymm} for a schematic of the action in Eq.\,(\ref{eq:fullaction}). This 
division is also convenient because the first two terms correspond to rather 
standard modules in the literature \cite{Hosotani:1983xw,Hosotani:1988bm,Hatanaka:1998yp,Antoniadis:2001cv,Cheng:2002iz,vonGersdorff:2002as} and \cite{Randall:1999ee} $+$ \cite{GW}.  (See Reference \cite{Sundrum:2005jf} for a review of both modules.) 
In the next three subsections we will flesh out 
each of these terms in the action. In this section we 
 focus on the physics that 
we actually {\it want}. Of course it is important to consider 
corrections that might destabilize the desired inflationary outcome, and 
we do this in Section\,\ref{sec:leadcorr}.

\subsection{The Inflaton Throat}
\label{sec:inflatoninterval}

	The inflaton throat is a finite interval whose size we denote by $\pi L$.  We realize this interval as an $S_1/\mathbb{Z}_2$ orbifold and we identify the orbifold fixed points at $y = 0~ $and$~ \pi L$. This will be a convenient 
way of assigning boundary conditions for fields.
The bulk fields of the throat consist of
 a $U(1)$ gauge field $A_M$, a heavy charged scalar $\Sigma$, 5D gravity, and neutral scalar fields that stabilize the  
throat geometry.  On top of all these fields
the throat will also contain very heavy fields associated with the UV completion of the non-renormalizable 5D effective field theory.  However, the physics
we will rely on will require propagation across the entire throat 
and will therefore be insensitive to the very short range effects of the 
UV completion.  In particular, the purpose of this sector is to realize a 4D 
inflaton as the $A_5$ component of a gauge field \cite{ArkaniHamed:2003mz,Kaplan:2003aj}.

  The physics of  the inflaton throat  is 
given by:
	\begin{eqnarray}
&S_{infl}& =  S_{U(1)} +S_{\Sigma}+ S_{stab} +
S_{gravity},\nonumber\\
&S_{U(1)}& = -\frac{1}{4}\int d^4x \int dy\sqrt{G}G^{MN}G^{KL}F_{MK}F_{NL}\nonumber\\
&S_{\Sigma}& = \int d^4x \int dy\sqrt{G}\left\{G^{MN}D_M\Sigma^{\dagger} 
D_N\Sigma-m_{\Sigma}^2|\Sigma|^2\right\}
\end{eqnarray}

	The extra-dimensional geometry of the inflaton throat can be 
stabilized by the Goldberger-Wise (GW) mechanism \cite{GW}.  This is the 
content of $S_{stab}$. For simplicity we
 consider an inflaton throat which is very mildly warped.\symbolfootnote[4]{In other words, the warp factor $e^{-k|y|}$ stays order 1 over the length of the throat.}  Stabilization energies can then be not much 
smaller than the compactification scale which we will take much higher 
than the scales relevant to inflation, so that the throat is 
essentially rigid on the scales of interest. That is 
we  can simply take the 5D metric in this throat to be of the form
\begin{equation}
ds^2 \approx g_{\mu \nu}(x) d x^{\mu}  d x^{\nu} -  dy^2,
\end{equation}
 $g_{\mu \nu}(x)$ is the 4D zero-mode gravitational field. 
We will be more explicit about Goldberger-Wise stabilization dynamics 
in the waterfall throat where it plays a more important role.

We focus more carefully on the gauge sector.
We assign orbifold parity to the gauge field in the following way: $P(y) = -y$, $P(A_{\mu}) = - A_{\mu}$, $P(A_5) = A_5$ and $\Sigma$ is 
 assigned even parity.  This parity assignment ensures the existence of an approximately massless zero mode for the fifth component of the gauge field: $A_5$.  Indeed the rest 
of $A_5$ can be gauged away, so that $A_5$ can be taken to be $y$-independent.
The 
remaining  modes are all Kaluza-Klein (KK) modes and 
have masses at the compactification scale $\sim 1/L$. $\Sigma$ does not
yield any light mode for large bulk mass.
Gauge-invariantly the $A_5$ zero-mode corresponds to the 
Wilson loop around the compactification,  
$W(x^{\mu}) \equiv \exp(ig_5 \oint_{S_1}dyA_5(x^{\mu}, y))$, 
which in the present gauge is
$W(x) = e^{i 2 \pi L g_5 A_5(x)}$. We see that the zero mode is an angular 
variable, essentially a 4D pseudo-Nambu-Goldstone boson (PNGB). 
The corresponding ``decay constant,'' $f_{\pi}$,
follows by going to the canonically 
normalized  $\Pi(x) \equiv  A_5(x) \sqrt{\pi L}$, 
\begin{equation}
S_{U(1)} \xrightarrow[IR]{} S_{\Pi} = \int \sqrt{-g}\,d^4x \frac{g^{\mu\,\nu}\partial_{\mu} \Pi\partial_{\nu} \Pi}{2},
\end{equation}
in terms of which the gauge-invariant observable is
\begin{equation}
W(x) \equiv e^{i \Pi(x)/f_{\pi}},
\end{equation}
with 
\begin{equation}
f_{\pi} = \frac{1}{2\,g_{5}\sqrt{\pi L}}.
\end{equation}

The  5D non-locality of $\Pi$ protects 
its potential from short-range effects such as the physics of the UV 
completion. 
Instead its potential is determined by the lightest charged 
fields that can propagate around the compactification, in the 
present case the 1-loop propagation of $\Sigma$, in a constant 
$\Pi$ background (constant 
since we are only after the $\Pi$ effective potential)\,\cite{Hosotani:1983xw,Hosotani:1988bm,Hatanaka:1998yp,Antoniadis:2001cv,Cheng:2002iz,vonGersdorff:2002as}. 
For $m_{\Sigma} \pi L \gg 1$ this is 
\begin{equation}
V_{\Sigma-loop}(\Pi) \sim -\frac{m_{\Sigma}^2}{8\pi^4 L^2}
e^{-2 \pi L m_{\Sigma}} \cos\left(\frac{\Pi}{\fpi}\right).
\label{eq:pioneloop}
\end{equation}
The exponential suppression is the Yukawa suppression for the 
massive field to virtually propagate around the compactification and 
``measure'' the Wilson loop $e^{i \Pi/f_{\pi}}$.  We will see that this potential can dominate the slow-roll of the inflaton during inflation.

Brane-localized couplings, predominantly linear or ``tadpole'' 
couplings of $\Sigma$, could affect the 
effective potential for $\Pi$.  They are allowed by the orbifold breaking of 
5D gauge invariance and we will make use of such couplings in subsection \ref{subsec:crosscoup}.

The various contributions to the 4D low-energy physics from this throat 
are then given by
\begin{equation}
\label{eq:infleff} 
{\cal L}_{eff, infl} = \sqrt{g} \left\{ M_5^3 \pi L {\cal R}^{(4)} + g^{\mu \nu} 
\frac{\partial_{\mu} \Pi \partial_{\nu} \Pi}{2} - V_{\Sigma,\,loop}(\Pi) - 
\rm constant \right\}, 
\end{equation}
where $g_{\mu \nu}(x)$ is the 4D gravity zero mode, ${\cal R}^{(4)}$ is its
curvature, $M_5$ is the 5D 
Planck scale, and we do not specify the constant piece of the potential 
since brane tensions effectively act as ``counterterms''
 that we can dial for it.
(We allow ourselves one fine-tuning in the end, namely the 4D cosmological 
constant after reheating.)

\subsection{The Waterfall Throat}
\label{sec:waterfallthroat}
The waterfall throat is a finite interval whose size we denote by $\pi r$.  We realize this interval as an $S_1/\mathbb{Z}_2$ orbifold, distinct from that of the inflaton throat.  We identify the orbifold fixed points at $y = 0~ $and$~ \pi r$ as the UV and IR branes of this highly warped throat, where the UV brane 
is the intermediate brane of the entire set-up. 
Here, we limit ourselves to the physics relevant after reheating, in 
particular describing the true vacuum of the theory, deferring the 
subtler structure and fields needed for the inflationary era until 
subsection \ref{subsec:crosscoup}. For our present purposes the waterfall throat is just a copy 
of the RS1 model (leaving out the Standard Model (SM) fields for simplicity), 
with minimal Goldberger-Wise stabilization \cite{GW}. In particular, the field content 
we start with consists just of the 5D metric and the 5D Goldberger-Wise 
scalar $\Phi$. The structure of these is well-known and we will not write it out 
explicitly. 

Instead we merely recall that the light 4D degrees of freedom 
consist of the 4D gravitational zero mode $g_{\mu \nu}(x)$, 
 (the same as in subsection \ref{sec:inflatoninterval}) and the  
 4D radion of the waterfall throat, contributing to the 4D effective theory,
\begin{equation}
\label{eq:watereff}
{\cal L}_{eff,water} = \sqrt{g} \left\{ \frac{M_5^3}{k} (1 - \omega^2){\cal R}^{(4)}  + \frac{6\,M_5^3}{k} 
g^{\mu \nu} \partial_{\mu} \omega \partial_{\nu} \omega + \lambda \frac{\omega^4}{4}
-  \kappa \frac{\omega^{4 + \gamma}}{4+\gamma}
- {\rm constant} \right\},
\end{equation}
where the maximal warp factor 
$\omega(x) \equiv e^{- k \pi r(x)}$ is the convenient choice of radion field, 
$r(x)$ is the dynamical proper length of the throat, and $1/k$ is the 
 radius of curvature of the approximately $AdS_5$ local bulk geometry,
\begin{equation}
ds^2 \approx e^{-2ky} g_{\mu \nu}(x) dx^{\mu} dx^{\nu} - d y^2.
\end{equation}
The $\kappa$ coupling is the leading effect of integrating out the 5D 
Goldberger-Wise scalar with $m_{\Phi}^2 \equiv \gamma (4 + \gamma) k^2$ and with 
brane-localized couplings $J_{IR}\Phi$ and $J_{UV}\Phi$:
\begin{equation}
\frac{\kappa}{4+\G} \sim \frac{J_{IR}J_{UV}}{k}.
\end{equation}  
The constant contribution is again not specified because the UV brane tension 
acts as a counterterm that allows us to choose it to cancel the 
cosmological constant after reheating. The $\lambda$ coupling arises 
when the IR boundary ``tension'' is not at the RS1 tuned value. We assume this 
more generic ``de-tuned'' choice of tension, 
\begin{equation}
\frac{\lambda}{4} = T_{RS} - T.
\end{equation}

The $\lambda, \kappa$ couplings (chosen positive) lead to a stabilizing 
potential for  $\omega$ and hence a finite length of throat. This 
vacuum state will describe the vacuum of the theory {\it after reheating}. 
In the next subsection we will turn to the fields, symmetries 
and physics that put us in an inflationary state. 

Again, we have left out the Standard Model fields for ease of explanation, but we note that the simplest possibility is to place the Standard Model on the IR brane of the waterfall throat.  The radion, as the dual to the dilaton, couples to breaking of scale invariance, i.e. mass terms.  If the visible sector contains a scalar $\phi_{vis}$ with mass $m_{vis}$ then the radion coupling to the visible sector is given by:
\begin{equation}
\label{eq:radioncoupling}
{\cal L}_{\w-vis} \approx \delta\w\,m_{vis}^2 \phi_{vis}^2,
\end{equation}
where $\delta\w$ is the fluctuation of the radion field from its vev.  The lesson here is that the radion will decay preferentially to the heaviest particle kinematically accessible.  This might be a SM field, or more generally, a heavier field on the IR brane with appreciable couplings and decays to the SM.

\subsection{The Inflaton-Waterfall Cross-Coupling}
\label{subsec:crosscoup}

We now add two new bulk (orbifold-even) real scalars, $\chi_{i = 1,2}$,
 to the warped waterfall throat (but {\it not} in the inflaton throat), 
which however have tachyonic masses, 
\begin{equation}
S_{tachyons} = \int d^4 x \int dy \sqrt{G}\,\frac{1}{2}\left\{ 
G^{MN}  \partial_M\chi_i \partial_N\chi_{i}- m_i^2 \chi_i^2 \right\}, ~ ~ 0 > m_i^2 
\geq - 4 k^2.
\end{equation} 
While tachyons above the 
Breitenlohner-Freedman bound \cite{Breitenlohner:1982bm} $m_5^2 \geq -4 k^2$ do not imply any 
instability in infinite $AdS_5$ spacetime, they are a potential source of 
instability for RS1. UV-brane localized tadpole (linear) couplings in the 
tachyon field will source growing profiles in the IR which can blow up before
reaching the IR boundary.  We therefore introduce two new discrete symmetries,
$\Zed_2, \Zed_2'$, to control tadpole couplings.
We take $\chi_1$ to be odd under $\Zed_2$ and even under $\Zed_2'$, and 
 $\chi_2$ to be odd under $\Zed_2'$ and even under $\Zed_2$.  We will also assign $\Sigma$, from the inflaton throat, to be odd under $\Zed_2'$ and even under $\Zed_2$. We will take 
these symmetries to be fully broken on the boundaries of the inflaton and 
waterfall throats. We only need the protection of these symmetries in the bulk 
and on the UV (middle) brane; $\Zed_2'$ is taken to be exact in these regions, while $\Zed_2$ is slightly broken.

Finally, we add a set of localized couplings on all three branes and boundaries 
of our entire set-up. Denoting the brane induced 4-metric by $G_{\mu \nu}^{ind}$,
\begin{eqnarray}
 S_{infl\,bdry} &=& \int d^4 x \sqrt{G_{ind}}\,\left.\left(\vphantom{\sqrt{G_{ind}}}J_1\Sigma + J_1^{\dagger}\Sigma^{\dagger}\right)
\right|_{boundary}
\nonumber \\
 S_{water\,bdry} &=& \int d^4 x \left.\sqrt{G_{ind}} \left(\vphantom{\sqrt{G_{ind}}}J_4 \chi_2 -
 J_5 \chi_1 - \mu_1\,\chi_1^2 - \mu_2\,\chi_2^2\right)\right|_{boundary} 
\nonumber \\
 S_{mid\,brane} &=&  \int d^4 x \sqrt{G_{ind}} \left\{\vphantom{\sqrt{G_{ind}}}
j_3 \chi_1 + 
  \left.(J_2\,\Sigma + J_2^{\dagger}\,\Sigma^{\dagger}) \chi_2 - \mu_1^{(\textrm{UV)}}\chi_1^2 - \mu_2^{(\textrm{UV})}\chi_2^2  \right\}\right|_{brane},\nonumber\\
\end{eqnarray}
 where ``$|_{brane,\,boundary}$'' just means the evaluation of the 
bulk field at the $y$-location of the brane or boundary in question. 
The $J,j$'s and $\mu$'s are constants. 

We denote the $\chi_1$ source 
on the middle brane with a small $j$ to indicate that it is taken to be 
much smaller than the fundamental scale, whereas the capital $J$ coefficients 
are supposed to be only modestly smaller than the fundamental scale. 
The smallness of $j_3$ is technically natural, corresponding to a 
small breaking of the $\Zed_2$ symmetry on this brane. The $J_2$ coupling is linear in 
each of $\Sigma$ and $\chi_2$, a kind of brane-localized mixing mass term, 
coupling a waterfall throat field to an inflaton throat field at their shared 
border. Note that this coupling is fully symmetric, but separate tadpole couplings in either $\Sigma$ or $\chi_2$ are forbidden on the middle brane.

\begin{figure}[ht]
\centering
\includegraphics[scale=0.4]{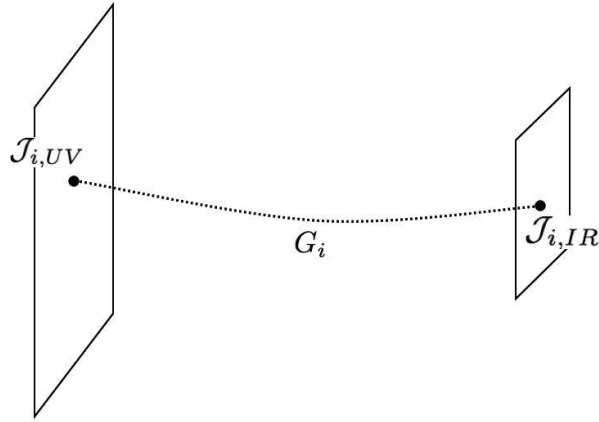}
\caption{Generic mixed position-momentum space Feynman diagram giving rise to the potential in Eq.\,(\ref{eq:w2plusa1}).  The index $i$ runs over the bulk fields: $\chi_1$, $\chi_2$ and $\Phi$.  The functions $G_i$ are the mixed position-momentum space propagators appropriate to the bulk fields and the sources $\mathcal{J}_{i,UV}$ and $\mathcal{J}_{i,IR}$ are the tadpoles: $j_3$, $J_2$, $J_{UV}$ and $J_5$, $J_4$, $J_{IR}$ respectively.}
\label{fig:generalwaterfall}
\end{figure}

Assembling the above pieces defines
\begin{equation}
S_{cross} \equiv S_{tachyons} + S_{infl\,bdry} + S_{water\,bdry} + 
S_{mid\,brane}.
\end{equation}
Let us see what leading corrections this makes to our 4D effective theory.
We will choose the brane/boundary tachyon masses, $\mu$,  to all be positive and order $k$ so that neither
tachyon field leads to any 4D light or 4D tachyonic modes in the KK decomposition. We can therefore
completely integrate out $\chi_i$ in arriving at the 4D effective theory \cite{Davoudiasl:2005uu}.
We begin by integrating out the $\chi_1$ tachyon, sourced by 
$j_3, J_5$, which then corrects the radion potential. We can do the Feynman 
diagrams shown in Fig.\,\ref{fig:generalwaterfall} in mixed position-momentum space, where the 
4D momentum is zero since we are implicitly computing an effective 
 potential correction in the radion background (that is, an $x$-independent 
IR brane position). Using the RS scalar field propagator \cite{Contino:2003ve} with brane masses $\sim k$, we find\symbolfootnote[2]{For compactness we are absorbing an $\mathcal{O}(1)$,\;$\alpha$-dependent coefficient into the definition of the tadpole couplings.} 
\begin{equation}
\label{eq:w2plusa1}
\Delta V_{eff}(\Pi, \omega) = -\frac{j_3 J_5}{k} 
\omega^{2+ \alpha_1}.
\end{equation}

\begin{figure}[ht]
\centering
\includegraphics[scale=0.5]{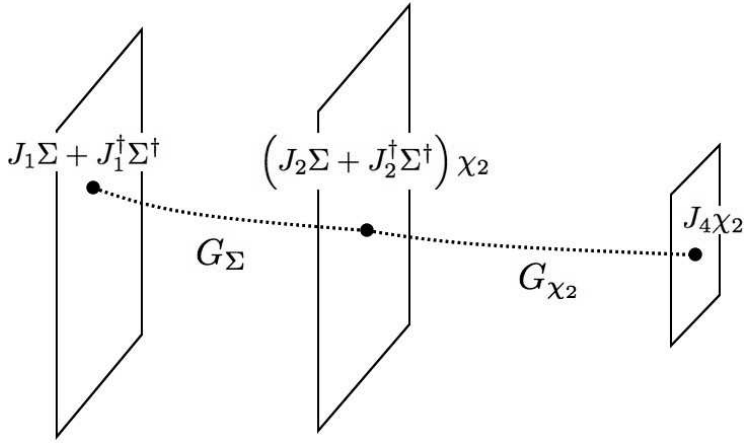}
\caption{Feynman diagram giving rise to the `waterfall' coupling in Eq.\,(\ref{eq:Vwaterfall}).  The 4D coupling is naturally small because it arises from dynamics which are non-local in the extra dimension.}
\label{fig:waterfallcoupling}
\end{figure}

Similarly, we can do the diagram of Fig.\,\ref{fig:waterfallcoupling} by integrating out $\chi_2$ and 
$\Sigma$. We get the product of their propagators $\times$ sources, 
\begin{equation}
\label{eq:Vwaterfall}
\Delta V_{eff}(\Pi, \omega) = -\frac{J_1 J_2 J_4}{m_{\Sigma}k}
 e^{-m_{\Sigma} \pi L} \cos\left(\frac{\Pi}{2\fpi}+\sigma\right)\, \omega^{2+ \alpha_2},
\end{equation}
where $\sigma = \arg(J_1\,J_2).$
The $\Pi$-dependence arises because we are computing the $\Sigma$ propagator
in a constant $A_5$ background, in order to determined the latter's effective 
potential. We therefore end up with the parallel transport phase factor across
the inflaton throat for $\Sigma$, plus that for $\Sigma^{\dagger}$. 
The factor of $1/2$ in the cosine relative to Eq.\,(\ref{eq:pioneloop}) arises because we 
are only propagating one-way across the inflaton throat rather than round-trip.

Note that the fundamental scale near the IR boundary of the waterfall throat 
is warped down $\propto \langle \omega \rangle$, as is standard in RS1. That translates into 
maximal energy densities localized near the IR boundary that scale 
$\propto \langle \omega \rangle^4$. Given that the radion itself, $\omega(x)$, is localized 
near the IR boundary, we see that the tachyonic nature of the $m_{\chi}^2 < 
0,  \alpha < 2$, translates into potential energy densities that drop more 
slowly  with small $\langle \omega \rangle$ 
(which will be the regime of interest). If all other factors were 
near the fundamental scale, the potential energy densities would violate 
 the maximal energy density bound of 5D effective field theory. This 
 reflects  the general 
danger of  instability that tachyonic scalars pose to RS1. 
This is 
precisely where the protective symmetries $\Zed_2, \Zed_{2}'$ come in. 
The small breaking of $\Zed_2$ on the middle brane translates into a small 
$j_3$ which can suppress the first potential energy density in Eq.\,(\ref{eq:w2plusa1}). 
The exact $\Zed_{2}'$ on the middle brane requires the linear $\chi_2$
dependence to multiply $\Sigma$. This in turn requires a $\Sigma$ 
propagation across the inflaton throat which is Yukawa-suppressed by 
$e^{-\pi L m_{\Sigma}}$. We will ensure that we choose parameters such that
these suppression factors keep us within effective field theory control. 

\section{The 4D $\Pi-\omega$ Effective Field Theory}
\label{sec:4Dtwofield}
Below the compactification scales of the two throats, there are just three 
4D effective fields, $\Pi(x), \omega(x)$, and the zero mode metric 
$g_{\mu \nu}(x)$.  Adding the contributions from Eqs.\,(\ref{eq:pioneloop},\,\ref{eq:infleff},\,\ref{eq:w2plusa1},\,\ref{eq:Vwaterfall} and \ref{eq:watereff}) the 4D effective 
theory is given by
\begin{eqnarray}
\label{eq:5dS4d}
{\cal L}_{eff} &=& \sqrt{g} \left\{ g^{\mu \nu}\frac{ \partial_{\mu} \Pi \partial_{\nu} \Pi}{2}
+ \frac{6\,M_5^3}{k} g^{\mu \nu} \partial_{\mu} \omega \partial_{\nu} \omega 
+ \left(\frac{M_5^3}{k} (1 -\omega^2)  + M_5^3 \pi L\right) {\cal R}^{(4)}\right. 
\nonumber \\
&-& \left.\vphantom{g^{\mu \nu}\frac{ \partial_{\mu} \Pi \partial_{\nu} \Pi}{2}} V_{eff}(\omega, \Pi)\right\}\,, \nonumber \\
V_{eff}(\omega, \Pi) &=& -\frac{j_3J_5}{k}\w^{2 + \alpha_1} -  
\frac{J_1  J_2 J_4}{m_{\Sigma} k}
e^{-m_{\Sigma} \pi L}\cos\left(\frac{\Pi}{2\fpi}+\sigma\right) \w^{2+\alpha_2}-\frac{\lambda}{4}\w^4 + 
\kappa \frac{\w^{4+\gamma}}{4+\G} \nonumber \\ 
&-& \frac{m^2_{\Sigma}}{8 \pi^4 L^2} e^{-2 m_{\Sigma} \pi L} \cos \frac{\Pi}{f_{\pi}} + {\rm constant}.
\end{eqnarray}
The exponents are ordered as
\begin{equation}
0 <  \aon <  \atw < 2 < 2+\G
\end{equation} 
by our choice of bulk scalar masses in the waterfall throat. 

Although $\omega$ is a dynamical variable we will always remain in a 
regime where  $\omega \ll 1$. Therefore
the 4D effective Planck scale is simply given by: 
\begin{equation}
M_{Pl}^2 \approx M_5^3 (\pi L + 1/k).
\end{equation}


It is also convenient in what follows to trade  other 
fundamental parameters of our
 model appearing above for new constant parameters, 
$\delta, \omega_0, \omega_1, 
v_0, V_0$, implicitly defined by re-expressing the effective Lagrangian,
\begin{eqnarray}
\label{eq:s4d}
{\cal L}_{eff} &\approx& \sqrt{g} \left\{  g^{\mu \nu} \frac{\partial_{\mu} \Pi \partial_{\nu} \Pi}{2}
+ \frac{6\,M_5^3}{k} g^{\mu \nu} \partial_{\mu} \omega \partial_{\nu} \omega 
-  \frac{M_5^3}{k} \omega^2 {\cal R}^{(4)} + \frac{M_{Pl}^2}{2} {\cal R}^{(4)} \right.
\nonumber \\
&-& \left.\vphantom{\left\{  g^{\mu \nu} \frac{\partial_{\mu} \Pi \partial_{\nu} \Pi}{2}\right.}V_{eff}(\omega, \Pi)\right\}, \nonumber \\
V_{eff}(\omega, \Pi) &=& - \frac{\D \w_0^{2 - \aon}}{2+\alpha_1}\w^{2+\alpha_1} -  \frac{\D \w_0^{2 - \alpha_2}}{2+\alpha_2}\cos\left(\frac{\Pi}{2\fpi}+\sigma\right)\w^{2 + \alpha_2} - 
\kappa \frac{\w_1^{\G}}{4}\w^4+ \kappa \frac{\w^{4+\G}}{4+\G} +\nonumber\\
&-&v_0 \cosp +V_0.
\end{eqnarray}
It is also worthwhile summarizing here the constraints on field space.
The angular nature of $\Pi$ means that without loss of generality, 
$0 \leq \Pi(x) \leq 4 \pi \fpi$, and the warp factor nature of $\omega$ means
that $0 \leq \w(x) \leq 1$.

\begin{figure}[ht]
\centering
\includegraphics[scale=0.6]{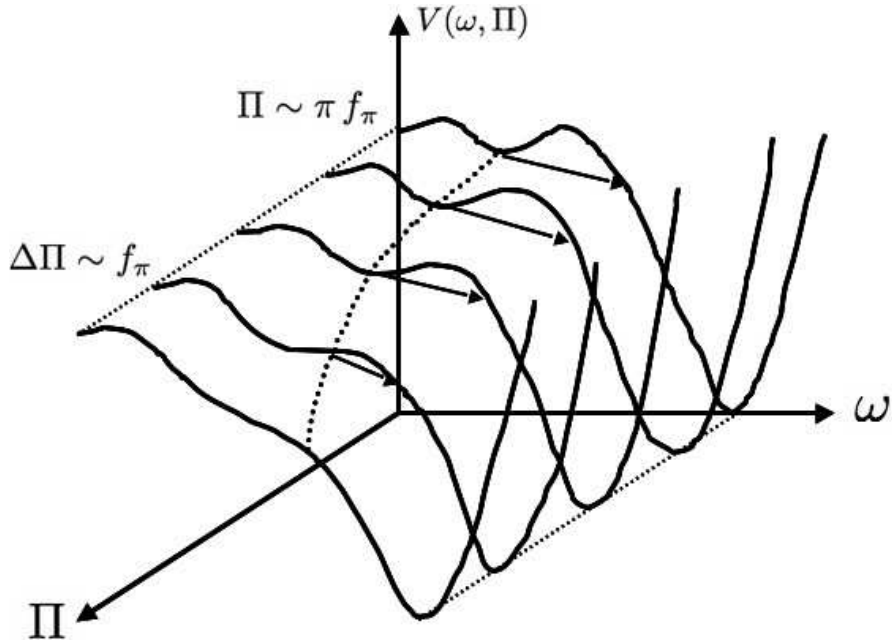}
\caption{Schematic of the two-field potential  - \textit{not to scale}.  Both fields move along classical trajectories determined by Eq.\,(\ref{eq:s4d}) as in the dotted path.  However, as indicated by the arrows in the diagram, $\w$ can tunnel through a steadily decreasing barrier and then roll to the global minimum thus ending inflation early in a portion of the universe. }
\label{fig:2field}
\end{figure}

We have chosen the above parametrization
 in anticipation of inflating at a metastable vacuum - $\w_{inflation} \simeq \w_0$ - and then rolling to reheat at 
the global minimum at $\w_{reheat} \simeq \w_1 > \omega_0$. 
We must 
ensure that while $\omega \sim \omega_0$, $\Pi$ is rolling slowly toward $0$ under the influence of the effective potential, resulting 
in cosmic inflation.  We will eventually choose a region of parameter space where the $v_0$ coupling dominates the slow-roll of $\Pi$.  Taking $\sigma \sim \pi/2$, the potential is metastable in  $\omega \sim \omega_0$ for $\Pi \sim \pi\fpi$ and becomes unstable with $\omega$ rapidly rolling to $\omega_1$ for sufficiently small $\Pi$, resulting in reheating.
 This is the hybrid inflation mechanism we intend to pursue below in our 
model. A schematic diagram of the potential as well as the path taken in field 
space is given in Fig.\,\ref{fig:2field}.

In the remainder of this section, we study the {\it classical}
waterfall field dynamics 
assuming that the inflaton $\Pi$ is indeed evolving very slowly. We study 
the inflationary regime where $\cos\left(\frac{\Pi}{2\fpi} + \sigma\right) \sim -1$, and then the 
regime after reheating, where $\cos\left(\frac{\Pi}{2\fpi} + \sigma\right) \sim 1$. This will allow us to pin down some 
of the constraints on our parameters. In the next section we consider the 
quantum subtlety of $\omega$ being able to tunnel from $\omega_0$ to 
$\omega_1$ even before it can classically roll there (again indicated in 
Fig.\,\ref{fig:2field}). 
In Section\,\ref{subsub:slowroll}, we ensure that the effective potential for $\Pi$ does indeed 
satisfy the constraints for slow-roll inflation. 

\subsection{The Effective Waterfall Potential for Fixed $\Pi$}

\begin{figure}[ht]
\centering
\includegraphics[scale=0.4]{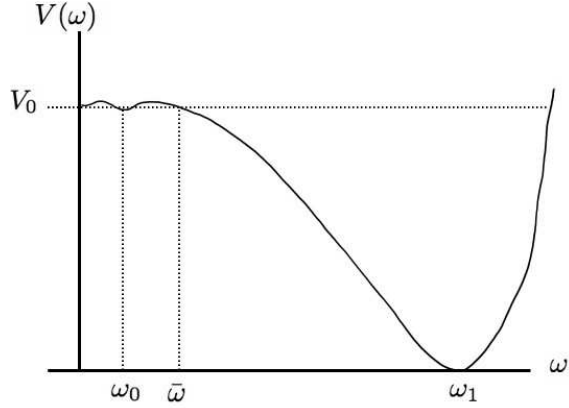}
\caption{Schematic of the waterfall potential during inflation - \textit{not to scale}.  The inflaton is near $\pi \fpi$ with $\sigma \sim \pi/2$, so that $\cos\left(\frac{\Pi}{2\fpi}+ \sigma\right) \approx -1$ and the waterfall sits in the metastable vacuum $\w_{inflation} \approx \w_0$ where the energy density is roughly $V_0$. }
\label{fig:potential}
\end{figure}

	During inflation we approximate $\cos\left(\frac{\Pi}{2\fpi} + \sigma\right) \sim -1$, with $\sigma \sim \pi/2$.
 Furthermore, if successfully inflating, 
the 4D Ricci scalar is well approximated by $\ric^{(4)} = 12\,H^2$ which contributes to the $\omega$ mass through the $\ric^{(4)}$ coupling in Eq.\,(\ref{eq:5dS4d}).  
Thus, we consider the following effective potential for the waterfall; see Fig.\,\ref{fig:potential} for a schematic of the waterfall potential during inflation:

\begin{equation}
\label{eq:watereff}
V_{eff}(\w) = 12\,H^2\frac{M_{5}^{3}}{k}\w^2 - \frac{\D \w_0^{2 - \aon}}{2+\aon}\w^{2+\aon} + \frac{\D\w_0^{2-\atw}}{2+\atw}\w^{2+\atw} - \kappa \frac{\w_1^{\G}}{4}\w^4+\kappa \frac{\w^{4+\G}}{4+\G} +   V_0.
\end{equation}

The simple operation of our model relies on the following constraints on the 
above potential:

\begin{itemize}

\item We have neglected the $\Pi$ potential energy density 
$v_0$ above by assuming  parameters such that:
\begin{equation}
\label{eq:neglectv0}
v_0 \ll V_0.
\end{equation}

\item We will arrange that 
during the evolution of the universe $\omega$ grows 
from $\sim \omega_0$ to $\sim \omega_1$. In this range it is important that 
the tachyon-induced potential energies never exceed the 
maximal energy densities allowed within 5D effective field theory
near the waterfall IR boundary,  as discussed 
at the end of subsection\,\ref{subsec:crosscoup}. Even more strongly, we wish to stay below 
the energies at which the KK modes can be excited, so that we 
always are within 4D effective field theory. Since the 
KK masses of the warped throat are of order $k \omega$, this is satisfied 
if our physical\symbolfootnote[1]{Note that our $\omega$ field is not canonically normalized, see the $\omega$ kinetic term in Eq.\,(\ref{eq:s4d}).} radion mass is smaller than $(k \omega)^2$. We will arrange this 
by choosing:
\begin{equation}
\label{eq:tachsafe}
\delta \ll 12\,\pi^2\,M_{5}^3\,k.
\end{equation}


\item The second and third terms  
create a local minimum at $\w_0$ as long as the effects of the first, 
fourth and fifth terms are negligible when $\w \sim \w_0$. This is satisfied 
if:
\begin{equation}
\label{eq:wheavyhubble}
12 \frac{M_5^3}{k} H^2 \ll \delta \omega_0^2,
\end{equation}
\begin{equation}
\label{eq:w0ind}
\kappa \omega_1^{\gamma} \ll \delta.
\end{equation}

\item The fourth and fifth terms produce a global minimum at $\w_1$ if 
we can neglect the first, second and third terms for $\w \sim \w_1$. 
We want  the ordering of minima to satisfy
\begin{equation}
\omega_1 \gg \omega_0.
\end{equation}
Then the first three terms can be neglected for $\omega \sim \omega_1$ if
\begin{equation}
\label{eq:w1indw0}
12\frac{M_5^3}{k} H^2 \ll \kappa \omega_1^{2+\gamma},
\end{equation}
\begin{equation}
\label{eq:w1indw0two}
\delta \omega_0^{2- \alpha_2} \ll \kappa \omega_1^{2 + \gamma - \alpha_2}.
\end{equation}

\end{itemize}

At late times, $\Pi$ slow-rolls to smaller values and the local minimum in the $\omega$ direction 
moves from $\omega_0$  to larger values,  and then eventually 
disappears altogether.  At this point, 
$\omega$ rolls to the global $\omega_1$ minimum, corresponding to reheating.

\begin{itemize}

\item The energy density at the global minimum $\omega_1$
 is dominated by the last three terms. The 
vacuum energy here must vanish, requiring us to tune,
\begin{equation}
\label{eq:V0}
V_0 \approx \kappa \frac{\G\w_1^{4+\G}}{4(4+\G)}, 
\end{equation}
  This is the usual (fine-)tuning 
of the 4D cosmological constant of our universe to zero\symbolfootnote[3]{Note that this shows up in the 4D action as a tuning of the constant term in Eq.\,(\ref{eq:5dS4d}).}.

\item Given our earlier choices, $V_0$ dominates over all the other terms in 
Eq.\,(\ref{eq:watereff}) during inflation, when $\omega \simeq \omega_0$. Therefore, the inflationary 
Hubble constant is given by: 
\begin{equation}
H^2_{inflation} \approx \kappa \frac{\G\w_1^{4+\G}}{12(4+\G) M_{Pl}^2}.
\end{equation}  
\newpage

\item We leave the details of reheating after inflation to future work, but it is possible to estimate the reheat temperature in two simple cases.
\subparagraph{Instantaneous Decay}If the waterfall were to decay instantaneously into relativistic particle species, thereby converting all of the inflationary energy density into radiation, one can estimate the reheat temperature:
\begin{eqnarray}
T_{instantaneous}^4 \approx V_0 \approx \kappa \w_1^{4+\G}\nonumber\\
T_{instantaneous} \approx \kappa^{1/4} \w_1^{1+\nicefrac{\G}{4}}.
\end{eqnarray}
But, within the present framework the waterfall will be longer lived.  (With the addition of another brane in the vicinity of $\w_1$ instantaneous reheating can be accomplished via brane collision, as in \cite{Dvali:1998pa}.)  Therefore, $T_{instantaneous}$ represents the theoretically maximum possible reheat temperature.  It is smaller than the warped down (by $\omega_1$) KK scale if: 
\begin{equation}
\label{eq:reheat1}
\kappa \omega_1^{\gamma} \ll (\pi k)^4,
\end{equation}
ensuring that KK modes are not excited during reheating.

\subparagraph{Waterfall Decay} In the present framework, the waterfall does not decay instantaneously, but has some lifetime $\tau_{\w}\sim\Gamma_{\w}^{-1}$ set by couplings of the form in Eq.\,(\ref{eq:radioncoupling}).  As discussed above, the largest coupling (and hence the dominant decay mode) will be to the heaviest kinematically accessible degree of freedom.  In this scenario, there is a period between the end of inflation and the radiation dominated epoch during which the energy density in the universe is dominated by the coherent oscillations of the waterfall field.  In this case, the IR brane reheat temperature is given by:
\begin{equation}
\label{eq:reheat2}
T_{RH} \sim \sqrt{\Gamma_{\w}M_{Pl}}.
\end{equation}
See, for example, Ref.\,\cite{Kolb:1988aj}.
In Table \ref{table:Cosmo} below, we estimate an upper bound for $T_{RH}$ assuming the IR brane Lagrangian contains a field whose mass is $m_{vis} \lesssim m_{\w}/2$ which very rapidly decays into much lighter SM degrees of freedom.

\end{itemize} 

\section{Tunneling out of Inflation}
\label{sec:tunnel}

Fig.\,\ref{fig:2field} illustrates the classical path in field space corresponding to hybrid 
inflation in our model. But the figure also shows that quantum mechanically
there is the possibility of tunneling to the $\omega_1$ vacuum. And of course,
the fields can follow the classical path for some time and then tunnel 
in the $\omega$ direction to the true minimum. Indeed the further one rolls in 
the $\Pi$ direction, the lower the barrier to tunneling, increasing its 
probability. In any case, tunneling 
provides a more abrupt end to inflation, and must be taken into account.

When tunneling occurs a bubble of the true vacuum of the potential is 
nucleated, while outside it we are still in the inflationary phase.
  The size of this bubble is initially some microscopic scale determined 
by the potential and then the bubble will grow as the universe expands. 
Regions of space in which inflation ended by tunneling will lie within 
the remains of such a bubble, while regions of space in which 
 inflation proceeds predominantly classically will have standard features. 
The phenomenology of tunneling bubbles is certainly worthy of further study 
since they seem to be a generic issue in our class of models, 
and may give a new class of cosmological signatures in the CMB or large scale 
structure; see \cite{Liddle:1991tr,Marra:2007pm,Lavaux:2009wm,Blau:1986cw} and references therein for existing studies in similar directions. We will take this up in future work.
 But in the present paper
we will take a more conservative approach, by identifying the region of our 
parameter space in which there is only a small
 probablity that there are any bubbles in our Universe which are on 
the scales on which we observe the CMB. That is we identify a very safe 
region of parameter space in this first paper, at the expense of having 
a very standard phenomenology.

	Precision measurements of the CMB indicate that over observable scales - roughly 1 Mpc to $(\textrm{few}\times 10^3)$ Mpc - temperature fluctuations are very small: $\frac{\delta T}{T} \sim 10^{-5}$. Measurements on these scales translate 
into nearly Gaussian quantum fluctuations of the inflaton that froze out 
between the sixtieth to fiftieth e-foldings of standard inflation when they 
reached horizon size.  Tunneling however will lead to the nucleation of bubbles
of a microscopically determined critical size $R_b$ 
if this is smaller than $H^{-1}_{inflation}$. These  bubbles will then 
 grow with the expansion of the Universe.  We are concerned with bubble regions
 that will have grown large enough to provide observable features within the
CMB. Given $R_b < H^{-1}_{inflation}$ such observable bubbles must have been 
nucleated before the fiftieth e-folding of inflation. We will impose a 
constraint on parameter space that ensures small probability of having 
even one such bubble in our observable universe, again with the hope of 
relaxing this tight constraint in future work.

Let us denote the microscopically determined 
probability of nucleating a bubble per unit 
volume per unit time by $\Gamma$, which we will put bounds on 
further below. Given the exponential growth of spatial volume 
during inflation, most bubbles will be formed at the latest times, in our
case the time of the fiftieth e-folding, when the universe had a volume 
$\sim (e^{10} H^{-1}_{inflation})^3$. 
This last e-folding lasts for a period $\Delta t \sim H^{-1}_{inflation}$. 
Therefore the number of potentially observable bubbles in our universe is:
\begin{equation}
\label{eq:bubblenumber}
N_{bubbles} \approx e^{30} \frac{\Gamma}{H^4_{inflation}}.
\end{equation}
We want $N_{bubbles} < 1$ for maximal safety. 

\begin{figure}[b]
\centering
\includegraphics[scale=0.4]{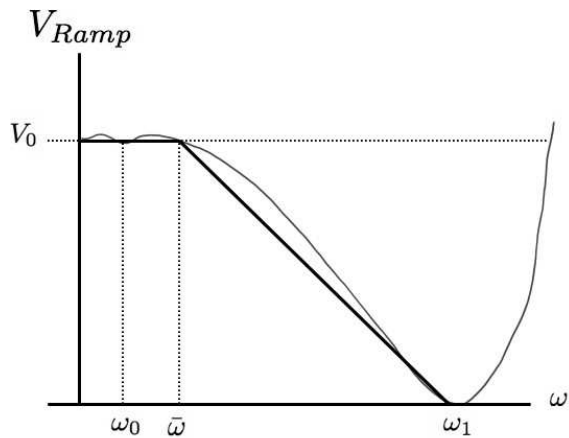}
\caption{The ramp approximation to the potential of Fig.\,\ref{fig:potential}, \textit{not to scale}.  This approximation is a conservative estimate of the true bounce action.}
\label{fig:vramp}
\end{figure}

Let us now bound $\Gamma$. The tunneling rate for a given potential takes the 
form in the semi-classical limit, 
\begin{equation}
\label{eq:tunnelrate}
\Gamma \approx Ae^{-S_B},
\end{equation}
where $S_B$ is a Euclidean ``bounce'' classical action for the potential and 
$A$ is a ratio of functional determinants \cite{Coleman:1977py,Coleman:1980aw}.
As illustrated in Fig.\,\ref{fig:vramp} we can replace our true $\omega$ potential 
(for $\Pi \sim \pi \fpi$ at the fiftieth e-folding,
if we manage to achieve the slow-roll conditions)  by a 
simpler ``ramp'' potential, which clearly has a larger tunneling 
rate since it neglects the metastable mass of $\omega$ about $\omega_0$ and 
overestimates the slope of the potential near the escape point $\bar{\omega}$.  That is 
\begin{equation}
\Gamma < \Gamma_{ramp}.
\end{equation}
Ref.\,\cite{Lee:1985uv} estimated: 
\begin{eqnarray}
A &\sim& \left(\frac{6\,M_{5}^{3}}{k}\right)^{2} \bar{\omega}^4 \nonumber \\
S_B &=& \frac{32 \pi^2}{3} \left(\frac{6 \,M_{5}^{3}}{k}\right)^{3/2} \frac{\bar{\omega}^3}{K}, 
\end{eqnarray}
where $K$ is the ramp slope and the factors of $6 M_5^3/k$ take into account the non-canonical normalization of $\omega$.

The escape point $\bar{\omega}$ is determined within our earlier 
approximations by 
cancellation of the third and fourth terms of Eq.\,(\ref{eq:watereff}):
\begin{equation}
\label{eq:defwbar}
\frac{\D  \w_0^{2 - \atw}}{2+\atw}\bar\w^{2+\atw}  \approx \kappa \frac{\w_1^{\G}}{4}\bar\w^4 \longrightarrow \bar\w \approx \w_0 \left(\frac{4 \D}{\kappa 
(2+\atw) \w_1^\G}\right)^{1/(2-\atw)}.
\end{equation}
Given Eq.\,(\ref{eq:w1indw0two}), we see that $\bar{\omega} \gg \omega_0$. We will therefore 
neglect $\omega_0$. The ramp slope is therefore given by: 
\begin{equation}
\label{eq:tunnelslope}
K \approx \frac{V_0}{\w_{1}-\bar{\omega}}\sqrt{\frac{k}{6\,M_{5}^{3}}},
\end{equation}
where the square root factor again accounts for the non-canonical normalization of $\omega$.

Assembling the pieces of our bound,
\begin{equation}
\label{eq:tunnelconstraint}
N_{bubbles} < e^{30} \left(\frac{6\,M_{5}^{3}}{k}\right)^{2}\frac{\bar{\omega}^4}{H_{inflation}^4} 
e^{- \frac{32 \pi^2}{3} \left(\frac{6\,M_{5}^{3}}{k}\right)^{2} \frac{\bar{\omega}^3\,(\w_{1}-\bar{\w})}{V_0}} < 1.
\end{equation}

\section{4D $\Pi$ Effective Field Theory}
\label{sec:onefield}

\subsection{Integrating Out The Waterfall}	

During inflation, when $\cos\left(\frac{\Pi}{2\fpi} + \sigma \right) \sim -1$, the physical radion mass-squared at its 
metastable point $\omega \sim \omega_0$ is of order $\delta\omega_0^2\,k/(12\,M_{5}^{3})$, 
since we have arranged that 
 the potential in this region is dominated by the second and third 
terms of Eq.\,(\ref{eq:watereff}).  By Eq.\,(\ref{eq:wheavyhubble}), it follows that the radion is 
heavier than $H_{inflation}$. 
  This justifies integrating $\w$ out during inflation 
and considering the simple one-field 
slow-roll potential for $\Pi$ and comparing it with the standard single-field
inflation requirements.

 The dominant effect of integrating out $\omega$  is classical and 
involves 
extremizing $V_{eff}(\Pi, \omega)$ with respect to $\omega$ for constant 
fixed $\Pi$, and then plugging this $\omega_{infl}(\Pi)$ back into the potential. 
During inflation the two-field potential is well approximated by:
\begin{equation}
V(\w,\Pi) \approx - \frac{\D \w_0^{2 - \aon}}{2+\aon}\w^{2+\aon} - \cos
\left(\frac{\Pi}{2 f_{\pi}}+\sigma\right) \frac{\D  \w_0^{2 - \atw}}{2+\atw}\w^{2+\atw} - v_0 
\cosp  + V_0.
\label{eq:twofieldinflation}
\end{equation}
Extremizing with respect to $\omega$,
\begin{eqnarray}
\label{eq:winflate}
\partial_{\w}V\left.\right|_{\w_{infl}} &\approx& - \D \w_0^{2 - \aon}\w_{infl}^{1 + \aon} - \cos\left(\frac{\Pi}{2 f_{\pi}} +\sigma\right)\D  \w_0^{2 - \atw}\w_{infl}^{1+\atw} = 0\nonumber\\
\w_{infl}&\approx& \w_0\left[-\cos
\left(\frac{\Pi}{2 f_{\pi}} + \sigma\right) \right]^{1/(\aon - \atw)}.
\end{eqnarray}
Plugging $\omega_{infl}$ back into Eq.\,(\ref{eq:twofieldinflation}), taking $\sigma \sim \pi/2$ and expanding to ${\cal O}(\Delta\Pi)^2$ (where $\Delta\Pi \equiv \pi\fpi - \Pi$)
we find the single-field effective potential:
\begin{equation}
\label{eq:VPi}
V_{eff,infl} \approx V_0 - \frac{m_{\Pi}^2}{2} (\Delta\Pi)^2,
\end{equation}
\begin{equation}
m_{\Pi}^2 \approx  - \left(\frac{v_{0}}{\fpi^{2}} +\frac{\delta\,\w_{0}^{4}}{4(2+\atw)\,\fpi^{2}}\right).
\label{eq:mpi}
\end{equation}
As mentioned earlier, we will choose parameters for which the first term in Eq.\,(\ref{eq:mpi}) dominates.  Very near the end of inflation, as the inflaton nears the critical point at which  the waterfall potential turns over, these approximations break down. 
 Nevertheless, they are very good during the first few e-foldings when the constraints from precision CMB measurements are most important.

\subsection{Constraining The Slow-roll Potential}
\label{subsub:slowroll}
In order to discuss constraints from precision cosmology, we begin by defining the usual potential slow-roll parameters \cite{Baumann:2009ds}:
\begin{eqnarray}
\label{eq:slowrolldef}
\epsilon &\equiv& \frac{M^{2}_{Pl}}{2}\left(\frac{V'}{V}\right)^2 \ll 1\nonumber\\
\eta &\equiv& M^{2}_{Pl}\frac{V''}{V} \ll 1.
\end{eqnarray}
In models described by Eq.\,(\ref{eq:VPi}) these are approximately:
\begin{eqnarray}
\label{eq:slowroll}
\epsilon(\Delta\Pi) &\approx& \frac{M_{Pl}^{2}}{2}\frac{m_{\pi}^4(\Delta\Pi)^2}{V_0^2}\nonumber\\
\eta &\approx& M_{Pl}^{2}\frac{m^2_{\pi}}{V_0} \approx -\frac{M_{Pl}^{2}}{V_{0}}
\left(\frac{v_{0}}{\fpi^{2}} + \frac{\delta\,\w_{0}^{4}}{4(2+\atw)\,\fpi^{2}}\right).
\end{eqnarray}
As is typical in hybrid models $\epsilon < \eta^{2} \ll \eta$, so that the 5-year WMAP measurement of the scalar spectral index indicates the following central value for $\eta$ \cite{Komatsu:2008hk}:
\begin{equation}
\eta = -0.02.
\end{equation} 
Measurements of the scale of primordial density perturbations imply the following constraint on the potential \cite{Komatsu:2008hk}:
\begin{equation}
\label{eq:COBEnorm}
\left.\frac{V}{\epsilon}\right|_{k_{0}}\approx 5.68\times10^{-7} M^{4}_{Pl},
\end{equation}
where the subscript $k_{0}$ indicates that the quantity is to be evaluated when the scale $k_{0} = 0.05\,\textrm{Mpc}^{-1}$ left the horizon.  Given the above approximations we can find the number of e-foldings before inflation ends as follows:
\begin{equation}
N(\Pi) \approx \int_{\Pi}^{\Pi_{end}} \frac{V(\bar\Pi)}{V'(\bar\Pi)}d\bar\Pi \approx  \frac{1}{\eta}\ln\left(\frac{\Delta\Pi_{end}}{\Delta\Pi}\right).
\end{equation}
\section{Leading Corrections}
\label{sec:leadcorr}
We have now worked through the desired behavior of our model, arriving 
at two-field and one-field effective descriptions with the 
delicate potentials needed for satisfactory inflation. But we must be 
careful that there are not uncalculated corrections that follow from 
our 5D action which could destabilize our story. The fact that the 
waterfall and inflaton fields are realized as moduli that are 
non-local from the 5D point of view means that they are insensitive to the 
short-range effects of whatever very massive physics UV completes our 
5D effective theory. Instead we must consider {\it finite} loop effects. 

The $\Pi$ potential is only sensitive to charged particles that traverse the 
inflaton throat and are sensitive to the associated Wilson loop observable.
But such effects are Yukawa suppressed for massive charged particles, 
by $e^{- \pi L m_5}$. We are assuming that $\Sigma$ is considerably lighter
than other charged fields at the cutoff of the effective description. 
 The cross-coupling with $\omega$ already makes use of a one-way virtual 
traversal of the inflaton throat ending on boundary/brane localized sources. 
The leading correction that then involves a virtual ``round-trip'' in the 
inflaton throat, Yukawa-suppressed by $e^{- 2 \pi L m_{\Sigma}}$,
 is given by the vacuum loop of $\Sigma$ in the $\Pi$ ($A_5$) background, 
since the diagram is unsuppressed by any further small couplings. But we 
have already estimated this and we will choose parameters so that it dominates the inflaton 
potential up to a constant 
 during inflation. Higher corrections will be parametrically smaller. 
Once the vacuum loop satisfies the slow-roll conditions as discussed in the 
previous section, they will remain satisfied with the higher corrections. 

In the waterfall throat, however, we have only worked at classical order 
thusfar. The
leading corrections to the $\omega$ potential will arise from vacuum loops of 
the bulk fields that involve virtual round-trips and thereby measure the 
radius of compactification parametrized by $\langle \omega \rangle$. 
In the AdS/CFT dual 
language, our bulk scalar fields correspond to single-trace operators of 
a large-N type conformal field theory, and the UV brane sources linear in 
these fields correspond to deformations of the underlying CFT by 
these single-trace operators. But regardless of the quantum numbers of 
these operators, the double-trace ``square'' of such operators 
cannot be forbidden by symmetry (except possibly supersymmetry which we do not consider). The bulk loop corrections in the waterfall throat capture 
the effects of these double-trace deformations. The most important 
double-trace operator in the IR is the one with the lowest scaling dimension.
In the leading large-N limit this is given by twice the scaling 
dimension of the associated single-trace operator. The lowest dimension 
is dual by AdS/CFT to the lowest mass-squared in the warped throat, namely 
the tachyon $\chi_1$. By (scaling) dimensional analysis from the CFT-side
we therefore know that $\chi_1$ loops will contribute an 
$\omega$-dependent term to the effective potential 
$\propto \omega^{2(2+\alpha_1)}$. 

From the AdS side we get more information. Dimensional analysis for the one-loop correction to the 4D
potential says that it must be set by $k^4$, since $k$ sets the bulk mass-squared
of the tachyon field (for order one $\alpha_1$), and the only other scale seen by the
small quantum fluctuations (about the classical vacuum profile of the tachyon field) are the
brane/boundary masses $\mu$, which we have also taken to be  order $k$. Finally, there should be a loop
factor $\sim 1/(16 \pi^2)$ multiplying the answer. 
 We thereby estimate a correction 
 \begin{equation} 
\Delta_{loop} V_{eff} 
\sim \frac{k^4}{16 \pi^2} \omega^{2(2+\alpha_1)}.
\end{equation}
This can be compared with the explicit calculations of Ref.\,\cite{Garriga:2002vf} for $\alpha_1 \ll 1$.
We would like this correction to be small enough not to upset our earlier 
analysis of the $\omega$ potential during inflation. This requires that it 
is small compared to the other potential contributions $\sim \kappa
\omega_1^{4 + \gamma}$ at $\omega \sim \omega_1$. That is, we require 
\begin{equation}
\label{eq:dubtrace}
\frac{k^4}{16 \pi^2} \ll \kappa \omega_1^{\gamma - 2 \alpha_1}.
\end{equation}

\section{A Sample Set of Parameters}
\label{sec:sample}
	In this section we collect all of the theoretical and observational constraints on our model.  We pick a representative, but not fine-tuned, point in parameter space that satisfies all of these constraints and present the corresponding predictions for cosmological observables.  The free parameters in the 4D two field potential are: $\D$, $\fpi$, $\w_0$, $\w_1$, $\aon$, $\atw$, $\G$, $v_0$, $\kappa$ and $\sigma$.  Table\,\ref{table:constraints} summarizes the nature of the various constraints on this set of parameters, highlights their location in preceding sections and indicates which parameters they constrain.  Table\,\ref{table:numbers} contains a sample set of choices for the 4D parameters as well as corresponding choices for 5D parameters.  Table\,\ref{table:Cosmo} lists the predictions for the CMB and gravitational wave spectrum; i.e. predictions for the inflationary energy density $V_0$, the Hubble constant during inflation $H$, the scalar spectral index $n_S$, the running of the scalar spectral index $\frac{dn_S}{d\ln k}$ and the scalar-to-tensor ratio $r$.
	
\begin{table}[h]
\centering
\begin{tabular}{||c|c||c|c|c|c|c|c|c|c|c||}
\hline
Nature of Constraint  	&  Equation Number(s)  &    $\D$  &  $\fpi$  &  $\w_0$  &  $\w_1$  &  $\aon$  &  $\atw$  &  $\G$  &  $\kappa$ &$v_0$\\
\hline
$V_0$ dominates  		&  \ref{eq:neglectv0} 	&    		 		&  		&  		&  \ding{52}  & 		   &  		  &  \ding{52} & \ding{52}  &\\
Separated minima  &  \ref{eq:wheavyhubble},\,\ref{eq:w0ind},\,\ref{eq:w1indw0}, \ref{eq:w1indw0two} & \ding{52}  &  &  \ding{52}  &   \ding{52}  &   &  \ding{52}  &  \ding{52}  & \ding{52}&\\
No tachyon blow-up 	&  \ref{eq:tachsafe}  		& 	 	    \ding{52}&  		&    		&  		&  		    &   	   & 		 &&\\
Tachyon-Loop Corrections& \ref{eq:dubtrace}		&		&		&		&\ding{52}&\ding{52}&		&\ding{52}&\ding{52}&	\\
Reheat temperature		&	\ref{eq:reheat1}, \ref{eq:reheat2}			&	 	   	     &		&		& \ding{52} &		&		&\ding{52}&\ding{52}&\\
Slow-roll parameters 	 &  \ref{eq:slowrolldef}, \ref{eq:slowroll}  &  \ding{52}  &  \ding{52}  &  \ding{52}  &  \ding{52}  &  &  \ding{52}  &  \ding{52}&\ding{52}&\ding{52}\\
Spectral index normalization  	&  \ref{eq:COBEnorm}  &   \ding{52}  &  \ding{52}  &  \ding{52}  & \ding{52} &  &  \ding{52}  &  \ding{52}&\ding{52}&\ding{52}\\
Tunneling suppression  	&  \ref{eq:tunnelconstraint}   & \ding{52}  &	 & \ding{52} & \ding{52} &  & \ding{52} & \ding{52} & \ding{52} &\\
\hline
\end{tabular}
\caption{Summary of theoretical and observational constraints on the parameters of this model.  A \ding{52} indicates that a given parameter is constrained by a particular constraint.}
\label{table:constraints}
\end{table}

\begin{table}[h]
\centering
\begin{tabular}{||c|c||c|c||}
\hline
4D Parameter		&	Sample Value			 & 		5D Parameter			&		Sample Value					\\
\hline
$\D$				&	$1.9\times10^{-3}$		&			$k$				&		$0.1$						\\
$\fpi$			&	$2.3\times10^{-2}$		& 			$M_5$			&		$2\,k$						\\
$\w_0$			&	$5.3 \times 10^{-5}$ 	& 		$\pi\,L$				&		$100$						\\
$\bar{\w}$			& 	 $2.4\times10^{-3}$		&		$J_{1}$				&		$2\,k^{5/2}$		\\
$\w_1$			&	$1.5 \times 10^{-2}$ 	& 		$J_{2}$				&		$(1.3)\,k$					\\
$\aon$ 			&	$0.5$ 				& 		$j_{3}$				&		$(2.9\times10^{-5})\,k^{5/2}$	\\
$\atw$			&	$0.6$  				&          $J_{4}$					&		$2\,k^{5/2}$					\\
$\G$				&	$0.2$ 				&		$J_{5}$				&		$(0.1)k^{5/2}$				\\
$v_{0}$			&	$8.3\times10^{-20}$		&	$J_{IR}$					&		$(0.3)\,k^{5/2}$									\\
$\kappa$			&	$3.2\times10^{-5}$		&		$J_{UV}$				&			$(0.3)\,k^{5/2}$						\\
$\sigma$			&	$\sim\pi/2$			&			$g_{5}$			&			$1/\sqrt{2\,k}$						\\
$m_{\w}^2$		&	$1.3\times10^{-9}$		&		$m_{\Sigma}$			&			$(1.3)\,k$						\\
				&						&			$\lambda$		&		$(0.14)\,k^4$		\\
				&						&		All $\mu$'s			&		$k$					\\
\hline
\end{tabular}
\caption{Sample set of 4D and 5D parameters with all quantities are expressed in $M_{Pl} = 1$ units.}
\label{table:numbers}
\end{table}

\begin{table}[h]
\centering
\begin{tabular}{||c||c||}
\hline
Observable				& 	Predicted Value 								\\
\hline
$V_0$					&	$8\times10^{-15}\,M_{Pl}^4$ or $(7.3\times10^{14}\,\textrm{GeV})^4$		\\
$H_{inflation}$				& 	$5.2\times10^{-8}\,M_{Pl}$ or $1.3\times10^{11}\,\textrm{GeV}$		\\
$T_{RH}$					&      $\lesssim 5\times10^{-6}\,M_{Pl}$ or 1$\times10^{13}\,\textrm{GeV}$\\
$n_S - 1$					&	$-0.04$										\\
$\frac{dn_S}{d \ln k}$		&	$7.7\times10^{-4}$								\\
 $r$						&	$1.6\times10^{-7}$								\\

\hline
\end{tabular}
\caption{Predictions for cosmological observables from the parameters in Table\,\ref{table:numbers}.}
\label{table:Cosmo}
\end{table}

\newpage

   
\section{Discussion}
\label{sec:disc}
We have realized four-dimensional slow-roll hybrid inflation 
as the long-wavelength limit of a
 controlled five-dimensional effective field theory, and 
 presented a viable sample set of parameters. The maximal scale 
we invoke in our 5D model is the reduced 5D Planck scale, $M_5$. 
With this choice, non-renormalizable 5D quantum gravity amplitudes become 
strongly coupled at $\sim (16 \pi^2)^{1/3} M_5$. Our fundamental 
5D parameters with 
positive mass dimension are chosen to be somewhat smaller than this so that
the theory is indeed weakly-coupled. The only other non-renormalizable 
coupling outside gravity is the 5D gauge coupling $g_5$. 5D gauge theory 
amplitudes become strongly coupled at $\sim 16 \pi^2/g_5^2$. Our choice 
$g_5^2 \sim 1/M_5$ means that all our gauge theory calculations are 
safely below this strong-coupling scale.  On the other hand $g_5$ is strong 
enough to satisfy the ``gravity as the weakest force'' conjecture of Ref.\,\cite{ArkaniHamed:2006dz}.

By 
construction our sample parameter set is not fine-tuned 
in the usual sense.
The related short-distance quantum
divergences in standard  hybrid inflation, discussed in the introduction, 
are physically cut off by extra-dimensional separations. 
Instead viable inflation in our model follows once one is in the right 
``ballpark'' of parameter space. However, given that parameter space is 
multi-dimensional  this in itself constitutes a mild type of tuning.
The exception of course  is the fine cancellation of the overall 4D 
cosmological constant, for which our model contains no mechanism.

Three specific mild tunings in our construction are worth noting.  One is the fact that
the $\alpha_i$ are very close, corresponding to a near-degeneracy of the two tachyon fields in the
waterfall throat. We have not explained this approximate degeneracy, but it seems clear that
it could originate from an approximate symmetry that exchanges the two fields. Such near degeneracy
is characteristic of the Goldberger-Wise mechanism in order to avoid having to introduce very
small parameters (on the fundamental scale) by hand. 

 There however remains one parameter which
still is notably small, without symmetry protection.
While most of the waterfall field terms in our effective potential, Eq.\,(\ref{eq:watereff}), 
arise from propagation across throats, and are naturally small for small 
enough sources, the $\omega^4$ term is exceptional. As discussed in Section 
III C it arises purely locally in 5D, 
from the IR boundary tension being de-tuned from the 
RS1 value of $24 M_5^3 k$. It cannot be too large without making it 
impossible to satisfy all constraints within effective field theory control.
A quick estimate of the technically natural size of this ``de-tuning'' of the 
tension is $\geq M_5^4/(16 \pi^2)$, if one considers loop renormalization of 
the tension cut off by a mass scale $\sim M_5$. Our 
sample parameter set satisfies this technical naturalness criterion 
 for the $\omega^4$ coefficient. But given that $24 M_5^3 k \gg 
M_5^4/(16 \pi^2)$ we should ask why the IR boundary tension 
is even close to the RS1 value. Furthermore, while we have realized 
high-scale inflation by our choice of parameters in this paper, 
it appears that 
low-scale inflation requires a much smaller $\omega^4$ coefficient, 
namely an IR boundary tension  even closer to its 
RS1 value. This would violate even technical naturalness. The simplest 
resolution  is to assume that supersymmetry is preserved
in the vicinity of the waterfall IR boundary. This in no way commits us to 
overall supersymmetry,
 which we know is broken in the inflationary phase. For example 
supersymmetry breaking may originate on the UV brane without impacting the 
IR boundary tension. We hope to study a supersymmetric version of our model 
in future work, with a focus on understanding to what extent low-scale inflation can take 
place naturally.

We also found that complete suppression of quantum tunneling presented a 
strong constraint in our search for a viable region of parameter space. 
This suggests that more typically in our scenario, standard inflation would
be accompanied by some regions of space that ended inflation by tunneling 
during the first few e-foldings. These may lead to significant deviations 
from scale-invariance in the CMB spectrum on angular scales accessible by 
future measurements. Again, we hope to return to a fuller treatment of 
this possibility in future work.  

\acknowledgments The authors are grateful to N. Arkani-Hamed, C. Bennett, D. E. Kaplan, B. Tweedie and J. Wacker for useful discussions.  Raman Sundrum is grateful to the University of Maryland Center for Fundamental Physics for its hospitality while portions of this work were being completed.  The authors are supported by the National Science Foundation grant NSF-PHY-0401513 and by the Johns Hopkins Theoretical Interdisciplinary Physics and Astrophysics Center.


\end{document}